\documentstyle[twocolumn,eqsecnum,aps,epsf]{revtex}

\begin{document} 

\title{Black Hole--Scalar Field Interactions in Spherical Symmetry } 

\author{R. L. Marsa}

\address{ Department of Physics and Astronomy \\
The University of Pittsburgh \\
Pittsburgh, PA }

\author{M. W. Choptuik} 

\address{ Center for Relativity \\ 
The University of Texas at Austin \\ 
Austin, TX } 

\date{July 15, 1996}

\maketitle 
\widetext

\begin{abstract} 
We examine the interactions of a black hole with a massless scalar field
using a coordinate system which extends ingoing Eddington-Finkelstein
coordinates to dynamic spherically symmetric-spacetimes.  We avoid problems
with the singularity by excising the region of the black hole interior to
the apparent horizon.  We use a second-order finite difference scheme to
solve the equations. The resulting program is stable and convergent and
will run forever without problems.  We are able to observe
quasi-normal ringing and power-law tails as well an interesting nonlinear
feature. 
\end{abstract} 


\narrowtext

\section{Introduction}
\label{sec:intro}
A longstanding goal of relativists has been the long-time numerical
evolution of a black hole spacetime.  Such an evolution is difficult
because of the physical singularity the spacetime contains.  Despite their
early promise, it has not sufficed to use slicings (choices of time
coordinate) which avoid the physical singularity. Invariably, such
coordinate systems develop coordinate singularities in the vicinity of the
event horizon and computationally, physical and coordinate singularities
are essentially equally pathological.  Many years ago, Unruh suggested that
it might help to consider evolution of only the exterior of a black hole. 
In fact, he argued, since the black hole interior is, by definition,
causally disconnected from the rest of the universe, evolution of events
within the horizon constituted wasted computational effort.  In the
necessary absence of exact information concerning the spatial location of
the event horizon at any instant during the evolution of given black hole
initial data, Unruh proposed that the apparent horizon be tracked and used
as an approximation to the true horizon. Thornburg developed these ideas,
first in the context of generating initial data for $n$ black holes (each
of which could have arbitrary momentum and spin)~\cite{thornburg1985}, and
then in a program of research for solving the vacuum axisymmetric Einstein
equations~\cite{thornburg1987,thornburg1993}. However, the first clearly
successful application of the black-hole excising technique in a dynamical
situation came with the work of Seidel and Suen who studied spherical
evolution of the vacuum (Schwarzschild) as well as a self-gravitating
massless scalar field~\cite{seidelsuen1992}.  Extensions of this work are
discussed in Anninos et. al.~\cite{anninos1995}.  

$$~~~~~~~~~~~~$$
\bigskip\bigskip

The Seidel and Suen paper
is also notable for the introduction of a general technique, termed
\emph{causal differencing}, which ensures that, independently of the
details of the coordinate system adopted, the difference scheme's numerical
domain of dependence contains the physical domain of dependence (i.e. is
causal).  Alcubierre and Schutz have used a similar but somewhat more
general technique which they call \emph{causal reconnection} to treat the
wave equation on an arbitrarily moving grid \cite{alcubierreschutz1994}. 
Scheel \emph{et. al.} have recently used black hole excising in a study of
gravitational collapse in Brans-Dicke gravity
\cite{scheel1995a,scheel1995b}.

In this paper, we again use the black hole excising technique to examine
the interactions of a black hole with a massless, minimally coupled scalar
field in spherical symmetry.  Unlike previous work, we use a \emph{null}
based slicing to get a coordinate system (ingoing Eddington-Finkelstein)
which fits naturally with black hole excision (for a Schwarzschild black
hole, all variables are static and non-singular (appropriately smooth)
everywhere on the solution domain, and perturbations about Schwarzschild
give perturbations of this behavior).  Further, we introduce a modification
to these coordinates which allows for easy tracking of the apparent
horizon.  We are able to get a second-order, convergent evolution scheme
which will stably evolve forever, and which shows the expected effects of
quasi-normal ringing and the power law decay of the scalar field.  We also
see some interesting coordinate and nonlinear effects.

The plan of the remainder of the paper is as follows.  In
Sec.~\ref{sec:mmief}, and following some unpublished previous
work~\cite{choptuikunruh1988,choptuik1989}, we define the
minimally-modified ingoing Eddington-Finkelstein (MMIEF) coordinate system
and derive equations of motion for the gravitational and scalar fields
which are specialized to this coordinate system.  This derivation is based
on detailed calculations of the Einstein-Klein-Gordon equations of motion
in a general spherically symmetric coordinate system~\cite{marsa1995} which
have been summarized in Appendix A.  In Sec.~\ref{sec:regularity}, we
examine the issue of regularity at the origin, $r=0$, with the principal
result that use of the coordinate system must be restricted to cases where
matter never reaches $r=0$. We follow with a detailed description of our
finite difference approximations in Sec.~\ref{sec:fde} and a discussion of
initial data in Sec.~\ref{sec:id}.  We discuss the convergence and
stability of our difference solutions in Sec.~\ref{sec:convstab}---the
evidence presented there suggests that the scheme can be used to carry out
arbitrarily long evolutions.  Various physical and coordinate effects which
have emerged from our studies are discussed in sections
Sec.~\ref{sec:mscale}--Sec.~\ref{sec:nonlin} and we end with some
concluding remarks. 

We note that the programs used to generate the results described below were
written in RNPL (Rapid Numerical Prototyping Language), a language designed
by the authors \cite{marsa1995}.  This language allows for the easy and
compact expression of time-dependent systems of partial differential
equations and the rapid trial of various finite-difference techniques for
solving them.

\section{Minimally-Modified Ingoing Eddington-Finkelstein Coordinates}
\label{sec:mmief}
\begin{figure}
\centerline{\epsfxsize=3in\epsfbox{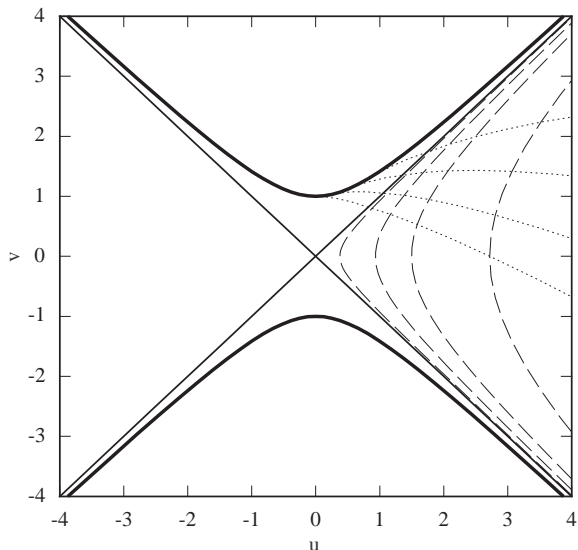}}
\caption{
The ingoing Eddington-Finkelstein slices in Kruskal-Szekeres coordinates.
The dotted lines are constant $t$, while the dashed lines are constant $r$.
The dark curves are the singularity.  The diagonal lines are the horizon ($r=2M$).
}
\label{fig:kruskal}
\end{figure}
We begin by recalling that the ingoing Eddington-Finkelstein (IEF)
coordinate system (see, for example, Chapter 30 of \cite{MTW1973}) is
defined for static, spherically-symmetric spacetimes (i.e. Schwarzschild)
and combines an areal (measures proper surface area) radial coordinate
$r$, with a time coordinate $t$, which is chosen so that the ingoing
tangent combination ${\vec \partial}_t - {\vec \partial}_r$ is null.  Fig.
\ref{fig:kruskal} shows some slices of constant IEF time plotted in
Kruskal-Szekeres coordinates.  Notice that all of the slices penetrate the
horizon and meet the singularity.

In generalizing IEF, we first consider the following general ``3+1'' form
for a time-dependent spherically-symmetric metric: 
\begin{equation}\label{eq:gmetric}
ds^2=\left(-\alpha^2+a^2\beta^2\right) dt^2 + 2 a^2\beta dt dr + a^2 dr^2 +
r^2b^2 d \Omega^2.
\end{equation}
Here, $a, b, \alpha$, and $\beta$ are functions of $r$ and $t$, and $d
\Omega^2$ is the metric on the unit-sphere.   We fix the spatial degree of
coordinate freedom by introducing a ``shifted'' areal coordinate $s$,
defined by $s\equiv r+f(t)$ for some as yet   undetermined function $f$. 
With this choice, the metric becomes \begin{equation}\label{eq:shiftmet} 
ds^2=\left(-\alpha^2+a^2\beta^2\right) dt^2 + 2 a^2\beta dt dr + a^2 dr^2 +
s^2 d \Omega^2. 
\end{equation} 
Comparison with the general form (\ref{eq:spmet}) yields the
identifications $s=rb$ and $b=1+f/r$.  From the general evolution equation
for $b$~(\ref{eq:spevb}), we have 
\begin{equation}
\dot{\left( rb \right)}=-\alpha rb K^{\theta}{}_{\theta} + \beta \left( rb
\right)'.
\end{equation}
Here and subsequently, overdots and primes denote partial differentiation
with respect to $t$ and $r$ respectively.  Solving this last equation for
the shift vector component (or simply ``the shift'') $\beta$, and noting
that $\dot{\left( rb \right)} = \dot{f}$, $\left( rb \right)'=1$, we obtain
the condition on the shift which must be dynamically enforced to keep the
metric function $s$ areal: \begin{equation}\label{eq:beta1}
\beta=\dot{f} + s\alpha K^{\theta}{}_{\theta} .
\end{equation}
We now fix the time slicing by demanding that the ingoing combination of
timelike and radial tangent vectors---$\vec{\partial_t} -
\vec{\partial_r}$---be null, exactly as in the case of the original IEF
coordinates.  This gives a condition on the metric, namely $g_{tt} -
2g_{tr} + g_{rr} =0$.  Using (\ref{eq:shiftmet}) this implies 
\begin{equation}
\alpha = \pm a \left( 1-\beta \right) .
\end{equation}
We choose the sign so $\alpha$ is positive for ${ \beta } \le 1$, that is
$\alpha = a \left( 1-\beta \right)$. Using this and (\ref{eq:beta1}) we get
\begin{equation}\label{eq:betadef}
\beta = \frac{\dot{f} + s a  K^{\theta}{}_{\theta}}{1+sa
K^{\theta}{}_{\theta}}
\qquad \alpha=\frac{a \left( 1 - \dot{f} \right)}{1 + s a 
K^{\theta}{}_{\theta}}.
\end{equation}
Hence the metric takes the form
\begin{equation}\label{eq:mmiefmet}
ds^2=a^2 \left( 2\beta - 1 \right) dt^2 + 2a^2\beta dtdr +
a^2dr^2+s^2d\Omega^2.
\end{equation}
Factoring the first three terms yields
\begin{equation}\label{eq:factmet}
ds^2=a^2\left( \left( 2\beta-1\right) dt + dr\right) \left( dt + dr \right)
+ s^2d\Omega^2,
\end{equation}
which shows that the characteristic speeds are
\begin{equation}\label{eq:charspeed}
c=-1,1-2\beta.
\end{equation}

We now specialize the general spherically-symmetric equations of
motion~(\ref{eq:speva})-(\ref{eq:spmom}) given in Appendix A to the MMIEF
coordinate system.  The full set of constraint and evolution equations for
the geometric variables is: 
\begin{eqnarray}\label{eq:ham}
a' + \frac{1}{2s}\left( a^3-a \right) &+& \frac{a^3s}{2}
K^{\theta}{}_{\theta}\left( 2K^r{}_r+ K^{\theta}{}_{\theta}\right) 
\nonumber \\ &-& 2\pi sa\left(\Phi^2+\Pi^2\right) = 0
\end{eqnarray}
\begin{equation}\label{eq:mom}
 K^{\theta}{}_{\theta}' + \frac{ K^{\theta}{}_{\theta} - K^r{}_r}{s} -
\frac{4\pi\Phi\Pi}{a}=0
\end{equation}
\begin{equation}\label{eq:aev}
\dot{a}=-a^2\left( 1-\beta\right) K^r{}_r +\left( a\beta\right)'
\end{equation}
\begin{eqnarray}\label{eq:kttev}
\dot{ K^{\theta}{}_{\theta}} = \beta K^{\theta}{}_{\theta}' &+& a\left(
1-\beta\right) K^{\theta}{}_{\theta}\left( K^r{}_r+2
K^{\theta}{}_{\theta}\right) \nonumber \\ &+& \frac{1-\beta}{s^2}\left(
a-\frac{1}{a}\right)+\frac{\beta'}{as}
\end{eqnarray}
\begin{eqnarray}\label{eq:krrev}
\dot{ K^r{}_r} &=& \beta K^r{}_r' + a\left( 1-\beta\right) K^r{}_r\left(
K^r{}_r+2 K^{\theta}{}_{\theta}\right) \nonumber \\ &+&
\frac{\beta-1}{a}\left[\frac{a''}{a}-\left(\frac{a'}{a}\right)^2-
\frac{2a'}{sa} + 8\pi\Phi^2\right] \nonumber \\ &+& \frac{\beta' a'}{a^2} +
\frac{\beta''}{a}
\end{eqnarray}
while the wave equation for the massless scalar field becomes the
first-order (in time) system:
\begin{equation}\label{eq:phiev}
\dot{\Phi}=\left(\beta\Phi+\left( 1-\beta\right)\Pi\right)'
\end{equation}
\begin{equation}\label{eq:piev}
\dot{\Pi}=\frac{1}{s^2}\left[ s^2\left(\beta\Pi+\left(
1-\beta\right)\Phi\right)\right]'
- \frac{2\dot{s}}{s}\Pi.
\end{equation}
We note that we have used the slicing condition, $\alpha=a(1 - \beta)$, to
eliminate the lapse function from the above set. In addition, as is always
the case in general relativistic dynamics, we have more constraint and
evolution equations governing the geometric variables than variables
themselves.  We adopt the often-used (and somewhat {\em ad hoc}) approach
of discretizing some sufficient subset of the equations with the
expectation that, provided the difference scheme converges, residuals of
discretized forms of the remaining equations will be of the same order in
the mesh spacing as the truncation error of the difference scheme itself. 
In fact, we have made considerable use of our freedom to construct schemes
based on various combinations of constraint and evolution equations in the
development of the stable second-order methods described in
Sec~\ref{sec:fde}.  It is entirely  possible that we could construct a
stable scheme without explicit use of the constraints but our attempts to
do so with the type of differencing described in Sec~\ref{sec:fde} were not
successful.

Since the function $f(t)$ (recall $s(r,t)\equiv r + f(t)$) is still
unspecified, we require one more equation to determine the time-evolution
of our model system.  We derive an evolution equation for $f$ by demanding
that a certain radial coordinate, $r=r_h$, be tied to the location of a
marginally trapped surface.  In general we will want to track the
\emph{outermost} marginally trapped surface, so in the following we will
assume that the surface we are tracking is, in fact, an apparent horizon.
We recall that if $S^\mu$ is an outward-pointing, space-like unit normal to
a marginally trapped surface, then it satisfies \cite{choptuik1986}
\begin{equation}\label{eq:aphor}
D_i S^i - K + S^i S^j K_{ij} = 0,
\end{equation}
where $D_i$ is the covariant derivative compatible with the 3-metric and
$K\equiv K^i{}_i$.  In spherical symmetry, and with the functional forms of
the 3-metric and extrinsic curvature given in Appendix A, this equation
reduces to
\begin{equation}\label{eq:spaphor}
\left( rb\right)' - arb K^{\theta}{}_{\theta} = 0.
\end{equation}
In the MMIEF system this is simply
\begin{equation}\label{eq:mmiefaphor}
as K^{\theta}{}_{\theta}=1.
\end{equation}
To keep the apparent horizon at fixed $r$, we demand that $\dot{\left( as
K^{\theta}{}_{\theta}\right)}|_{r_h} = 0$, where $r_h$ is the initial
position of the apparent horizon.  Solving this equation for $\dot{f}$ and
using the constraint and evolution equations to eliminate time and space
derivatives of the geometric variables, we find:
\begin{equation}\label{eq:fdot}
\dot{f} = \frac{4\pi s^2 \left( \Phi+\Pi\right)^2}{a^2} \bigg |_{r_h}.
\end{equation}

We are most interested in the situation where the initial data for our
spacetime describes a black hole of mass $M$ which is well separated from
any scalar field.   In this case we have $r_h = 2M$ and, provided
that~(\ref{eq:fdot}) is satisfied, the apparent horizon will remain at
$r=r_h=2M$.   The area of the apparent horizon, however, is given by $4\pi
s_h^2 \equiv 4\pi (r_h + f)^2$, and hence will increase as matter falls
into the black hole, in accord with physical expectations.

It is useful to write down the Schwarzschild solution in IEF coordinates. 
The metric is usually written as \cite{MTW1973}
\begin{equation}\label{eq:efmetnc}
ds^2 = -\left( 1-\frac{2M}{r} \right) d\tilde{V}^2 + 2d\tilde{V}dr +
r^2d\Omega^2,
\end{equation}
where $\tilde{V}$ is a null coordinate.  Defining a time-like coordinate $t
\equiv \tilde{V}-r$, the metric becomes:
\begin{eqnarray}\label{eq:efmettl}
ds^2 = &-& \left( 1-\frac{2M}{r}\right) dt^2 + \frac{4M}{r} dtdr \nonumber
\\ &+& \left(1 + \frac{2M}{r}\right) dr^2 + r^2 d\Omega^2.
\end{eqnarray}
Comparison of this last result with the general form~(\ref{eq:spmet})
yields the following expressions for the various metric components:
\begin{equation}\label{eq:schlapse}
{\bar \alpha} = \sqrt{\frac{r}{r+2M}} \qquad
{\bar \beta}=\frac{2M}{r+2M} \qquad
{\bar a}=\sqrt{\frac{r+2M}{r}},
\end{equation}
where we use overbars on quantities to stress that the expressions are
valid only for the vacuum case.  Using the above and equations
(\ref{eq:speva}) and (\ref{eq:spevb}), we can compute the extrinsic
curvature components:
\begin{equation}\label{eq:schktt}
 {\bar K^{\theta}{}_{\theta}}=\frac{2M\left( r+2M\right)}{\left( r\left(r +
2M\right)\right)^\frac{3}{2}} \qquad
{\bar  K^r{}_r}=\frac{-2M\left( r+M\right)}{\left( r\left(r +
2M\right)\right)^\frac{3}{2}}.
\end{equation}

Since we are working in spherical symmetry, we can meaningfully define the
mass (or mass aspect) function $m(r,t)$ which, at least in a vacuum region,
provides an invariant measure of the gravitational mass contained within
radius $r$ at time $t$.  Moreover, even when matter is present, $m(r,t)$
and $m'(r,t)$ are useful diagnostic quantities in our calculations.  The
mass in MMIEF coordinates can be computed from the surface area using the
general expression: 
\begin{equation}\label{eq:mass1}
m \left( r,t \right) = \frac{1}{2} s \left( 1 - \left( 16\pi A \right)^{-1}
A^{,\mu}A_{,\mu}\right),
\end{equation}
where $A=4\pi s^2$.  Substituting this expression for the area into
(\ref{eq:mass1}) and differentiating, we get 
\begin{equation}\label{eq:mass2}
m \left( r,t \right) = \frac{1}{2} s \left( 1 - \frac{1-\left( sa
K^{\theta}{}_{\theta}\right)^2}{a^2}\right).
\end{equation}
By making use of the evolution and constraint equations, we can write this
mass as an integral over the mass-density.  In this form we have
\begin{equation}\label{eq:mass3}
m\left( r,t \right) = \frac{s_h}{2} + 4\pi\int^r_{r_h}s^2\left(\frac{\Phi^2
+ \Pi^2}{2a^2} + s K^{\theta}{}_{\theta}\frac{\Phi\Pi}{a}\right) dr,
\end{equation}
where as before,  $r_h$ is the location of the apparent horizon and 
$s_h\equiv s(r_h)$.  We note here that as used above, $M$ represents the
mass of the black hole, that is $s_h/2$.  We also define $M_\infty$ as the
total mass in the spacetime, namely
\begin{eqnarray}\label{eq:minf}
M_\infty &\equiv& m\left(\infty\right) = M \nonumber \\ &+&
4\pi\int^\infty_{r_h} s^2\left(\frac{\Phi^2+\Pi^2}{2a^2} + s
K^{\theta}{}_{\theta}\frac{\Phi\Pi}{a}\right) dr.
\end{eqnarray}
Given that our finite difference grid does not extend to infinity, we will
approximate $M_\infty$ by taking the upper limit of the above integral to
be the outer boundary of  our computational domain.

\section{Regularity at the Origin}
\label{sec:regularity}
In cases where a black hole is not initially present or there is
insufficient mass in the scalar field to form a black hole via collapse,
the infalling matter will encounter the coordinate origin, $r=0$.  As is
generically the case when using spherical coordinates, the various
geometric and matter variables must satisfy regularity conditions as
$r\to0$ in order that the origin remain a regular point in the spacetime
(see \cite{bardeenpiran1983} for an extensive discussion of regularity
conditions).

Since $\phi$ is a scalar and $a$,$ K^{\theta}{}_{\theta}$, and $ K^r{}_r$
are components of rank-two tensors, we assume that they are even in $r$
near the origin. This means that their spatial derivatives must vanish at
$r=0$: 
\begin{equation}\label{eq:phiprime}
\phi'=a'= K^{\theta}{}_{\theta}'= K^r{}_r'=0.
\end{equation}

Elementary flatness near the origin in MMIEF coordinates dictates that
$a\left( 0,t\right)=1$.  An examination of the momentum constraint
(\ref{eq:mom}) shows that $K^{\theta}{}_{\theta}(0,t)= K^r{}_r$(0,t).  We
can find further conditions by examining the potentially divergent terms of
the evolution equation for $K^{\theta}{}_{\theta}$
(\ref{eq:kttev}).  These terms, which are those with powers of $r$ in the
denominator, can be collected and written as
\begin{equation}\label{eq:divterms}
\frac{\left( 1-\beta\right)\left( a^2-1\right) + r\beta'}{r^2a}.
\end{equation}
Clearly, as $r\to 0$ both the numerator and the denominator of
(\ref{eq:divterms}) go to zero, so we use L'H\^opital's rule to compute
the limit.  The derivative of the numerator is $2\left( 1-\beta\right) aa'
- \beta'\left( a^2 - 1\right)+\beta'+r\beta''$.  As $r\to 0$ this goes to
$\beta'$.  The derivative of the denominator is $r^2a'+2ra$.  Clearly this
goes to zero as $r$ goes to zero.  Thus, we must have $\beta(0,t)'=0$.

Since we still have an indeterminate form, we apply L'H\^opital's rule
again.  The second derivative of the numerator is $\left(
1-\beta\right)\left( 2aa''+2\left(a'\right)^2\right) - \beta''\left(
a^2-1\right) - 2a\beta'a'+2\beta''+r\beta'''$.  As $r$ goes to zero, this
goes to $2\left( 1-\beta\right) a''+2\beta''$.  The second derivative of
the denominator is $r^2a''+4ra'+2a$.  The limit of this is $2$.  Thus, we
have
\begin{equation}\label{eq:lim}
\lim_{r\to 0} \frac{\left( 1-\beta\right)\left( a^2-1\right) +
r\beta'}{r^2a} = \left( 1-\beta\right) a'' + \beta''.
\end{equation}

Let us now consider the behavior of the spatial derivatives of $\beta$ as
$r\to0$.  When there is no black hole present, the shift is given by
\begin{equation}\label{eq:nbshift}
\beta=\frac{ra K^{\theta}{}_{\theta}}{1+ra K^{\theta}{}_{\theta}}.
\end{equation}
From this we can easily see that $\beta$ is zero at the origin.  Now
\begin{equation}\label{eq:nbbetap}
\beta'=\frac{ra K^{\theta}{}_{\theta}'+ra' K^{\theta}{}_{\theta}+a
K^{\theta}{}_{\theta}}{\left( 1+ra K^{\theta}{}_{\theta}\right)^2}.
\end{equation}
Thus,
\begin{equation}\label{eq:limbetaprime}
\lim_{r\to 0}\beta'= K^{\theta}{}_{\theta}
\end{equation}
Since we have argued that $\beta'(0,t)=0$ and $K^{\theta}{}_{\theta}(0,t) =
K^r{}_r(0,t)$, we have
\begin{equation}
K^{\theta}{}_{\theta}(0,t) = K^r{}_r(0,t) = 0.
\end{equation}
Now the second derivative of $\beta$ is
\begin{eqnarray}\label{eq:betadprime}
\beta'' &=& \frac{ra K^{\theta}{}_{\theta}'' +2a K^{\theta}{}_{\theta}'
+2ra' K^{\theta}{}_{\theta}' +2a' K^{\theta}{}_{\theta} +ra''
K^{\theta}{}_{\theta}}{\left( 1 + ra K^{\theta}{}_{\theta}\right)^2}
\nonumber \\ &-& \frac{2\left(\left( ra
K^{\theta}{}_{\theta}\right)'\right)^2}{\left( 1+ra
K^{\theta}{}_{\theta}\right)^3}.
\end{eqnarray}
As $r\to0$, this expression vanishes.  Thus, $\beta''\left( 0,t\right) =0$.

Since $ K^{\theta}{}_{\theta}(0,t) = 0$, we must have ${\dot
K^{\theta}{}_{\theta}}(0,t) = 0$ and consequently the right hand side of
(\ref{eq:kttev}) must vanish as $r\to0$.  This will happen only if the
limit of (\ref{eq:divterms}) is zero, which, given the results deduced
above, can only happen if $a(0,t)''=0$. 

The evolution of the scalar field is accomplished through two auxiliary
variables, $\Phi$ and $\Pi$.  These are defined by (\ref{eq:defPhi}).  The
condition on $\Phi$ has already been stated in~(\ref{eq:phiprime})
\begin{equation}\label{eq:phicond}
\Phi\left( 0,t\right)=0,
\end{equation}
and it is easy to show that $\Pi$ must be even in $r$ near the origin. 
Thus we have 
\begin{equation}\label{eq:piprime}
\Pi'(0,t) = 0.
\end{equation}

\begin{table}
\caption{\label{tab:fdop}Two-Level Finite Difference Operators}
\begin{tabular}{lcr}
Operator & Definition & Expansion \\
\tableline
$\Delta^f_r f^n_i$ & $\left( -3f^n_i + 4f^n_{i+1}-f^n_{i+2}\right) /
2\Delta r$ &
$\partial_r f \big\vert^n_i + O\left(\Delta r^2\right)$ \\
$\Delta^b_r f^n_i$ & $\left( 3f^n_i - 4f^n_{i-1} + f^n_{i-2}\right)
/2\Delta r$ &
$\partial_r f \big\vert^n_i + O\left(\Delta r^2\right)$ \\
$\Delta_r f^n_i$ & $\left( f^n_{i+1}-f^n_{i-1}\right) /2\Delta r$ & 
$\partial_r f \big\vert^n_i + O\left(\Delta r^2\right)$ \\
$\Delta_t f^n_i$ & $\left( f^{n+1}_i-f^n_i\right)/\Delta t$ &
$\partial_t f \big\vert^{n+\frac{1}{2}}_i + O\left( \Delta t^2\right)$ \\
$\Delta^d_t f^n_i$ & $\left( f^{n+1}_i-f^n_i\right)/\Delta t + $ &
$\partial_t f \big\vert^{n+\frac{1}{2}}_i + O\left( \Delta t^2\right)$ \\
 & $\epsilon_{dis}[ 6f^n_i + f^n_{i-2}+f^n_{i+2} - $ & \\
 & $4\left( f^n_{i-1}+f^n_{i+1}\right)] /16\Delta t$ & \\
$\mu_t f^n_i$ & $\left( f^{n+1}_i + f^n_i\right) / 2$ &
$f\big\vert^{n+\frac{1}{2}}_i + O\left( \Delta t^2\right)$ \\
$\mu_r f^n_i$ & $\left( f^n_i+f^n_{i-1}\right) /2$ &
$f\big\vert^n_{i-\frac{1}{2}} + O\left( \Delta r^2\right)$ \\
$\Delta^{fa}_r f^n_i$ & $\mu_t\Delta^f_r f^n_i$ & $\partial_r f 
\big\vert^{n+\frac{1}{2}}_i + \qquad $ \\
 & & $O\left(\Delta r^2+\Delta t^2\right)$ \\
$\Delta^{ba}_r f^n_i$ & $\mu_t\Delta^b_r f^n_i$ & $\partial_r f
\big\vert^{n+\frac{1}{2}}_i + \qquad $ \\
 & & $O\left(\Delta r^2+\Delta t^2\right)$ \\
$\Delta^a_r f^n_i$ & $\mu_t\Delta_r f^n_i$ & $\partial_r f  
\big\vert^{n+\frac{1}{2}}_i + \qquad $ \\
 & & $O\left(\Delta r^2+\Delta t^2\right)$ \\
$\Delta^s_r f^n_i$ & $\left( f^{n+1}_i - f^{n+1}_{i-1} + \right.$ & 
$\partial_r f \big\vert^{n+\frac{1}{2}}_i + \qquad $ \\ 
 & $\left. f^n_{i+1}-f^n_i\right) /2 \Delta r $ & $O\left(\Delta r^2+\Delta t^2 \right. $ \\
 & & $ \left. + \Delta r \Delta t\right)$
\end{tabular}
\end{table}
Unfortunately, these regularity conditions are inconsistent with the
Hamiltonian constraint (\ref{eq:ham}).  If we solve~(\ref{eq:ham}) for
$a'$, take a radial derivative of the resulting expression, then take the
$r\to0$ limit, we find: 
\begin{equation}
a''(0,t) \propto \Pi(0,t)^2 \neq 0.
\end{equation}
Thus, we are lead to the conclusion that the MMIEF coordinate system will
admit no non-singular curvature at the origin. The only consistent
solutions near the origin describe flat space or a black hole.  Thus, MMIEF
is a ``good'' coordinate system to use only when a black hole already
exists in the spacetime.  In a spacetime without a black hole, the
equations will remain consistent as long as no matter encounters the
origin.  This will be the case if the scalar field is outgoing or if it
collapses to form a black hole before it encounters the origin.  For a
collapse problem, we could start with another coordinate system and change
to MMIEF coordinates if an apparent horizon forms.  If no such horizon
forms, there is really no need for the special horizon tracking properties
of MMIEF coordinates anyway.  Though we have not implemented such a scheme,
this approach was used successfully in 
\cite{seidelsuen1992,anninos1995,scheel1995a,scheel1995b}.

\section{Finite Difference Equations}
\label{sec:fde}
We use equations (\ref{eq:aev}), (\ref{eq:kttev}), (\ref{eq:phiev}),
(\ref{eq:piev}), and (\ref{eq:fdot}) to evolve $a$, $
K^{\theta}{}_{\theta}$, $\Phi$, $\Pi$, and $f$; equation (\ref{eq:mom}) to
find $ K^r{}_r$; and equation (\ref{eq:betadef}) to find $\beta$.  We 
solve these equations using finite difference techniques on a uniform mesh
with spacings $\Delta r$ and $\Delta t = \lambda \Delta r$, where the
Courant factor $\lambda$, is held fixed when we change the basic scale of
discretization.

Table \ref{tab:fdop} shows the operators we use in the discretizations. 
Note that while the derivative operators take a lower precedence than the
arithmetic operators, that is $\Delta_r {f^n_i}^2 = \left( {f^n_{i+1}}^2 -
{f^n_{i-1}}^2\right) / \Delta r$, the time averaging operator takes a
higher precedence, that is $\mu_t {f^n_i}^2 = \left( \mu_t f^n_i\right)^2$
and $\mu_t \left( a^n_i b^n_i \right) = \mu_t a^n_i \mu_t b^n_i$.

In the interior, the finite difference equations are:
\begin{equation}\label{eq:fdain}
\Delta^d_t a^n_i = -\mu_t \left( a^2 \left( 1-\beta\right)\right)^n_i +
\Delta^s_r\left( a\beta\right)^n_i,
\end{equation}
\begin{eqnarray}\label{eq:fdkttin}
\Delta^d_t { K^{\theta}{}_{\theta}}^n_i = \mu_t\beta^n_i\Delta^s_r {
K^{\theta}{}_{\theta}}^n_i + \mu_t\left(\frac{1-\beta}{s^2}
\left( a - \frac{1}{a}\right)\right)^n_i \nonumber \\ + \frac{\Delta^a_r
\beta^n_i}{\mu_t\left( as\right)^n_i} +
\mu_t\left( a\left( 1-\beta\right) K^{\theta}{}_{\theta}\left( 2
K^{\theta}{}_{\theta} +  K^r{}_r\right)\right)^n_i,
\end{eqnarray}
\begin{equation}\label{eq:fdkrrin}
\mu_t\left( \Delta_r K^{\theta}{}_{\theta}+\frac{ K^{\theta}{}_{\theta}-
K^r{}_r}{s}
- 4\pi\frac{\Phi \Pi}{a}\right)^n_i = 0,
\end{equation}
\begin{equation}\label{eq:fdphiin}
\Delta^d_t\Phi^n_i = \Delta^s_r\left(\beta\Phi + \left(
1-\beta\right)\Pi\right)^n_i,
\end{equation}
\begin{eqnarray}\label{eq:fdpiin}
\Delta^d_t\Pi^n_i &=& \frac{1}{\mu_t\left( s^n_i\right)^2} \Delta^s_r\left(
s^2\left(\beta\Pi + \left(
1-\beta\right)\Phi\right)\right)^n_i \nonumber \\ &-& 2 \Delta_t s^n_i
\mu_t\left(\frac{\Phi}{s}\right)^n_i, 
\end{eqnarray}
\begin{equation}\label{eq:fdfin}
\Delta_t f^n_i = 4\pi
\mu_t{\left(\frac{s\left(\Phi+\Pi\right)}{a}\right)^n_i}^2,
\end{equation}
\begin{equation}\label{eq:fdsin}
s^{n+1}_i=r_i+f^{n+1}_i,
\end{equation}
\begin{equation}\label{eq:fdbetain}
\mu_t \beta^n_i = \frac{\Delta_t f^n_i + \mu_t\left( as
K^{\theta}{}_{\theta}\right)^n_i}{1+\mu_t\left( as
K^{\theta}{}_{\theta}\right)^n_i}.
\end{equation}
These equations are applied everywhere in the interior except at the two
points next to the boundary points.  At these points, we use the same
equations except the dissipative time derivatives ($\Delta^d_t$) are
replaced by regular time derivatives ($\Delta_t$), since the value at $i+2$
or $i-2$ is not available at these locations.  It is interesting to note
that all of the spatial derivatives are \emph{angled} ($\Delta^s_r$) except
for the derivative of $\beta$ in equation (\ref{eq:fdkttin}) and the
derivative of $ K^{\theta}{}_{\theta}$ in equation (\ref{eq:fdkrrin}).
Switching any of these derivatives from angled to non-angled or from
non-angled to angled results in an instability.  We don't have an
explanation for why this particular combination of derivatives works, but
the use of angled derivatives here was motivated by their successful
application in previous work by Choptuik (see \cite{choptuik1986} and
\cite{choptuik1991}).  This is a good example of a technique that was
perfected by ``experimental'' numerical analysis.

As discussed previously, the inner boundary of our computational domain is
fixed to the apparent horizon, $r=r_h$.  In accord with the causal
properties of the apparent horizon, function values on the horizon can be
advanced without use of any information defined at $r < r_h$.  Therefore,
at $r=r_h$ we use the same equations employed in the interior, except that
we replace centered differences with forward differences. For example,
equation (\ref{eq:fdain}) becomes 
\begin{equation}\label{eq:fdah}
\Delta_t a^n_i = - \mu_t\left( a^2\left( 1-\beta\right)\right)^n_i +
\Delta^{fa}_r\left( a\beta\right)^n_i.
\end{equation}

Since the computational grid does not extend to infinity.  We adopt
outgoing conditions at the outer boundary, that is, we assume that no
radiation will enter the grid from large $r$.  While this is not strictly
true (there will be curvature back-scattering from the outgoing pulse),
this assumption provides a reasonable computational solution.

For the scalar field variables $\Phi$ and $\Pi$, the outgoing conditions
come from the condition on $\phi$, namely $s\phi \sim F\left( s-ct\right)$,
with $c=1-2\beta$ being the speed of the outgoing waves.  This means that
\begin{eqnarray}
\dot{\Phi}+\left( 1-2\beta\right) \Phi ' &+& \frac{\dot{s} + 1 - 2\beta -
2s\beta'}{s}\Phi \nonumber \\ &-&
\frac{\dot{s}+1-2\beta+2s\beta'}{s^2}\phi=0 
\end{eqnarray}
and
\begin{equation}
\left( 1-\beta\right) \left(\Pi+\Phi\right) +
\frac{1-2\beta+\dot{s}}{s}\phi = 0.
\end{equation}
These equations are discretized as
\begin{eqnarray}
\Delta_t\Phi^n_i &+& \left( 1-2\mu_t\beta^n_i\right) \Delta^b_r\Phi^n_i
\nonumber \\ &+& \frac{\Delta_t s^n_i + 1-2\mu_t\beta^n_i
-2\mu_ts^n_i\Delta^b_r\beta^n_i}{\mu_ts^n_i} \mu_t\Phi^n_i \nonumber \\ 
&-& \frac{\Delta_ts^n_i+1-2\mu_t\beta^n_i+2\mu_ts^n_i\Delta^b_r\beta^n_i}{
\left(\mu_ts^n_i\right)^2}\mu_t\phi^n_i = 0
\end{eqnarray}
and
\begin{equation}
\mu_t\left[\left( 1-\beta\right)\left(\Phi+\Pi\right)\right]^n_i +
\frac{1-2\mu_t\beta^n_i+\Delta_ts^n_i}
{\mu_ts^n_i}\mu_t\phi^n_i = 0.
\end{equation}

We can get approximate conditions on $a$ and $ K^{\theta}{}_{\theta}$ from
their Schwarzschild forms (\ref{eq:schlapse}) and (\ref{eq:schktt}) with
the mass aspect (\ref{eq:mass3}) used in place of $M$, since in vacuum, or
in any region where the scalar field's self-gravitation is negligible, $a$
and $K^{\theta}{}_{\theta}$ should take on these forms. For large $s$,
asymptotic expansion of these expressions gives: 
\begin{equation}
a \sim 1 + \frac{m(r,t)}{s} + O\left( s^{-2}\right)
\end{equation}
and
\begin{equation}
 K^{\theta}{}_{\theta} \sim \frac{2m(r,t)}{s^2} + O\left( s^{-3}\right).
\end{equation}
Thus, at large $s$ we have, to leading order
\begin{equation}
s\left( a-1\right) \sim m(r,t)
\end{equation}
and
\begin{equation}
s^2 K^{\theta}{}_{\theta} \sim m(r,t).
\end{equation}
We now deduce how $m(r,t)$ behaves in the large $s$, weak-field limit. 
Since $a\to 1$ and $ K^{\theta}{}_{\theta}\to 0$ we have
\begin{equation}
m \sim 4\pi \int s^2\left(\Phi^2+\Pi^2\right) ds.
\end{equation}
From the condition on $\phi$ we can see that $\Phi \sim G\left( u\right) /
s$ and $\Pi \sim G\left( u\right) / s$, where $u\equiv s-ct$.  Thus
\begin{equation}
m \sim 8\pi\int G^2\left( u\right) du \sim H\left( u\right),
\end{equation}
that is, m is ``outgoing'' at large $s$.  Therefore we get the following
conditions for $a$ and $ K^{\theta}{}_{\theta}$:
\begin{equation}
s\left( a-1\right) \sim H\left( s-ct\right)
\end{equation}
and
\begin{equation}
s^2 K^{\theta}{}_{\theta} \sim H\left( s-ct\right).
\end{equation}
These are discretized as
\begin{equation}
\Delta_t\left( s\left( a-1\right)\right)^n_i + \left(
1-2\mu_t\beta^n_i\right)\Delta^{ba}_r\left( s\left(
a-1\right)\right)^n_i = 0
\end{equation}
and
\begin{equation}
\Delta_t\left( s^2 K^{\theta}{}_{\theta}\right)^n_i + \left(
1-2\mu_t\beta^n_i\right)\Delta^{ba}_r\left( s^2
K^{\theta}{}_{\theta}\right)^n_i=0.
\end{equation}

\begin{figure}
\centerline{\epsfxsize=3in\epsfbox{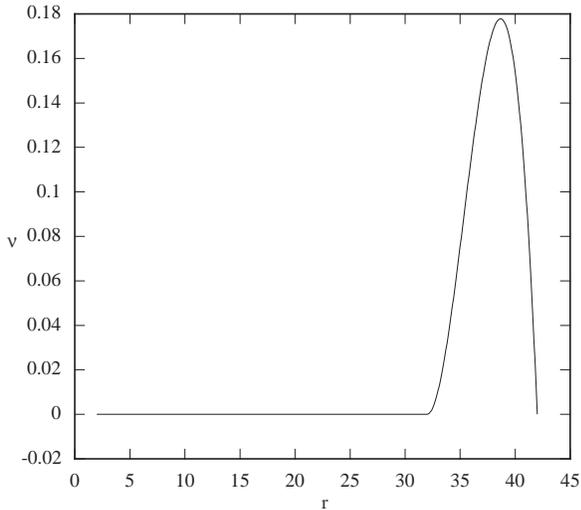}}
\caption{Sponge filter coefficient function for $A=1.0$ and $n=2$.}
\label{fig:sponge}
\end{figure}
The outgoing boundary condition reduces the amplitude of reflections off
the boundary, but unless the boundary is placed at very large $r$, these
reflections can still interfere with the results of a calculation.  To
further reduce the reflections, we use a \emph{sponge filter} as detailed
in \cite{choptuik1986}.  Briefly, this means that in the interior of the
grid, we add a term to the wave equation which effectively applies the
outgoing condition on a finite region rather than at a single radial
location.  For instance, we use the following modified evolution equation
for $\Phi$: 
\begin{eqnarray}
\dot{\Phi} &=& \left(\beta\Phi + \left( 1-\beta\right)\Pi\right)' \nonumber
\\ &-& \nu\left[\dot{\Phi} + \left(1-2\beta\right)\Phi' +
\frac{\dot{s}+1-2\beta-2s\beta'}{s}\Phi \right. \nonumber \\ &-& \left.
\frac{\dot{s} + 1-2\beta+2s\beta'}{s^2}\phi\right],
\end{eqnarray}
where the coefficient function, $\nu\left( r\right)$, is given by
$$
\nu(r) = A\frac{\left( r-r_s\right)^n \left( r_{max} - r\right)}{\left( r_{max}
-r_s\right)^{n+2}}\left( n+1\right)\left( n+2\right) 
$$
in the filtering region, $r_s \leq r \leq r_{max}$, and vanishes elsewhere.  
Here, $A$
and $n$ are parameters which can be adjusted to tune the filter.  Fig.
\ref{fig:sponge} shows $\nu$ for $A=1.0$ and $n=2$, the values used in this
work.

\section{Initial Data}
\label{sec:id}
We wish to examine the interactions of the black hole with compact ingoing
pulses of scalar field.  We can generate nearly ingoing pulses using the
following method.  Let $\phi \left ( r,t \right ) = F \left ( u \equiv r+t
\right ) / r $.  This gives $\dot{\phi}=\partial_u F / r$ and $\phi ' =
\partial_u F/r - F/r^2$.  For a compact pulse, we set $F$ to a Guassian of
the form
\begin{equation}\label{eq:guass}
F\left ( u \right) = A u^2 \exp \left( -\left( u-c \right)^d / \sigma^d
\right),
\end{equation}
where $d$ is an integer, and $c$ is the radial coordinate of the center of
the pulse.  This results in scalar field initial data of the form
\begin{equation}\label{eq:phiinit}
\phi=A r \exp\left( -\left( r-c\right)^d / \sigma^d\right)
\end{equation}
\begin{equation}\label{eq:Phiinit}
\Phi=\phi\left[\frac{1}{r}-\frac{d\left( r-c\right)^{d-1}}{\sigma^d}\right]
\end{equation}
\begin{equation}\label{eq:Piinit}
\Pi=\phi\left[\frac{2-\beta}{r\left( 1-\beta\right)}-\frac{d\left(
r-c\right)^{d-1}}{\sigma^d}\right] .
\end{equation}
We solve for $\beta,a$, and $K^\theta{}_\theta$ using equations
(\ref{eq:betadef}), (\ref{eq:ham}), and (\ref{eq:mom}).  Ostensibly,
$K^r{}_r(0,r)$ can be freely specified and the constraints can still be
satisfied by appropriate adjustments to the other geometric variables. 
However, an \emph{arbitrarily} chosen $K^r{}_r(0,r)$ combined with the
initial data for the scalar field given by
(\ref{eq:phiinit})--(\ref{eq:Piinit}) will not, in general, generate a
spacetime describing the desired physical scenario of a scalar pulse
initially infalling on a black hole.  We therefore adopt an ansatz for
$K^r{}_r(0,r)$ which is motivated by the observation that at the initial
time, the self-gravitation of the scalar field is generally relatively
weak, and thus the geometric variables should have approximately their
Schwarzschild form.  Specifically, we demand that $K^r{}_r(0,r)$ satisfy
the following equation (see (\ref{eq:schktt})):
\begin{equation}
K^r{}_r = \frac{-2m\left( r+m\right)}{\left( r\left(r +
2m\right)\right)^\frac{3}{2} },
\end{equation}
where $m$ is the mass aspect function defined by (\ref{eq:mass3}).  This
equation, along with the equation for the mass aspect (\ref{eq:mass3}) and
the constraints (\ref{eq:ham}) and (\ref{eq:mom}), is then solved
iteratively for any given initial scalar field configuration.  We have
found that initial data prepared in this manner \emph{does} generate
spacetimes of the type we seek even when the scalar field is significantly
self-gravitating at the initial time.

\section{Convergence and Stability}
\label{sec:convstab}
In order to assess the correctness and accuracy of our difference equations
and the program which solves them, we perform some tests. These include
computing convergence factors, performing a long-time vacuum
(Schwarzschild) evolution and comparing it to the known solution, and
performing a long-time strong-field evolution.

To measure convergence, we define the convergence factor for a grid
function $u$ by
\begin{equation}\label{eq:cf}
C_f\equiv\frac{|\hat{u}_{2h}-\hat{u}_{4h}|_2}{|\hat{u}_h-\hat{u}_{2h}|_2},
\end{equation}
where $\hat{u}_\alpha$ is a solution of the finite difference equations on
a grid with spacing $\alpha$, and the $\ell_2$ norm is defined in the usual
way, that is
\begin{equation}\label{eq:l2norm}
|\hat{u}|_2 \equiv \sqrt{\frac{\sum_{i=1}^N u_i^2}{N}}.
\end{equation}

\begin{figure}
\centerline{\epsfxsize=3in\epsfbox{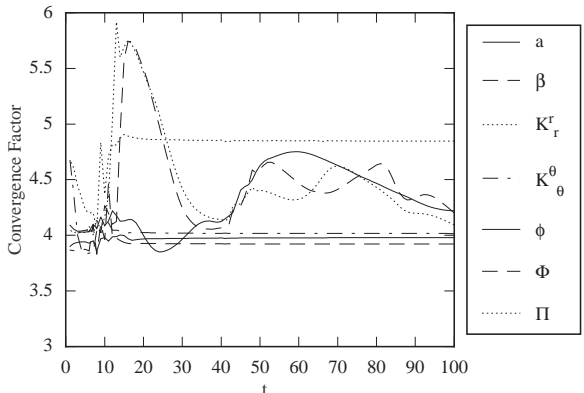}}
\caption{
Convergence factors as a function of time for a strong-field evolution with
$A=2.5\times 10^{-3}$, $c=10$, and $\sigma=2$.  These values are
approximately four, indicating second order convergence. 
}
\label{fig:conv}
\end{figure}
The convergence factors for the scalar field and geometric variables are
shown in Fig. \ref{fig:conv}.  These factors are computed from a
strong-field evolution.  They are approximately four throughout, indicating
second order convergence.

\begin{figure}
\centerline{\epsfxsize=3in\epsfbox{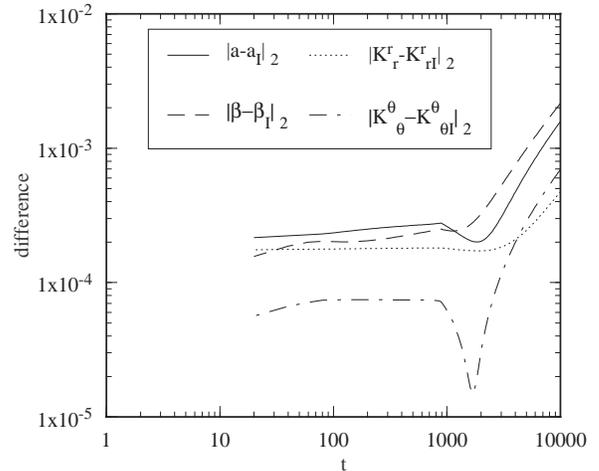}}
\caption{The $\ell_2$ norms of the differences of 
the computed vacuum geometrical
variables and their Schwarzschild values verses time.  Note that both axes
are logarithmic.  The outer boundary for this evolution was placed at
$r=82M$.
}
\label{fig:schdiff}
\end{figure}

Fig. \ref{fig:schdiff} shows the deviation of the evolved vacuum spacetime
from that of Schwarzschild.  We note however, that the spacetime being
evolved is not Schwarzschild, rather it is a black hole in a ``box'' with
rather \emph{ad hoc} boundary conditions.

\begin{figure}
\centerline{\epsfxsize=3in\epsfbox{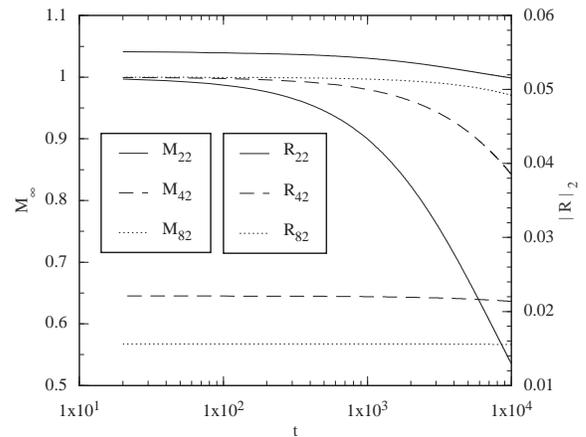}}
\caption{The mass and the $\ell_2$ norm of the 4-Ricci
scalar for vacuum spaces with varying outer boundary
position.  The subscripts indicate the position of the outer boundary in
units of $M$.  The time axis is logarithmic. }
\label{fig:riccimass}
\end{figure}
\begin{figure}
\centerline{\epsfxsize=3in\epsfbox{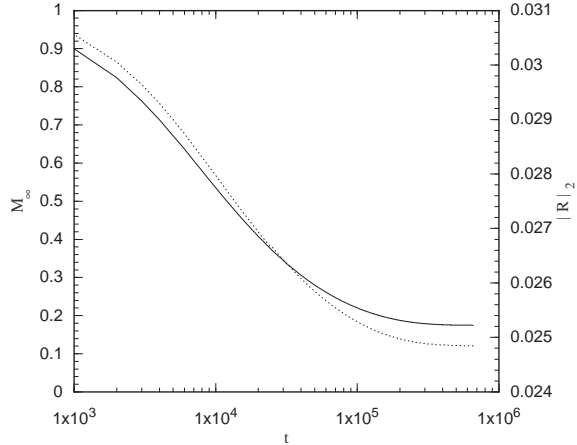}}
\caption{The mass (solid line) and the $\ell_2$ norm of the 4-Ricci scalar
(dotted line) for a vacuum evolution with outer boundary at $r=22M$.  Both
functions are asymptoting to a fixed value, indicating that the spacetime
is settling down to a static configuration.  The time axis is logarithmic.
}
\label{fig:longricci}
\end{figure}

At late times, the outer boundary condition causes the spacetime to
``drift'' away from the initial configuration.  Fig. \ref{fig:riccimass}
shows plots of the mass and the Ricci scalar for vacuum evolutions with
outer boundaries at $22M$, $42M$, and $82M$.  This plot clearly shows the
rather large effect the position of the outer boundary has on the late-time
evolution.  We could get better convergence of the difference solution to
the continuum solution (i.e. with boundary conditions only at spatial
infinity) by matching the interior Cauchy evolution to an exterior
characteristic evolution (see \cite{bishop1996,gomez1996}) or by using an
adaptive mesh refinement algorithm to push the the outer boundary to a
large radius without unduly increasing the computational load. 
Nevertheless, the position of the outer boundary has no effect on
stability.  This can be seen clearly in Fig. \ref{fig:longricci}.  The
curves in this graph are asymptoting to fixed values.  Thus,
although the computed spacetime drifts from Schwarzschild, it eventually
reaches a fixed configuration. 

\begin{figure}
\centerline{\epsfxsize=3in\epsfbox{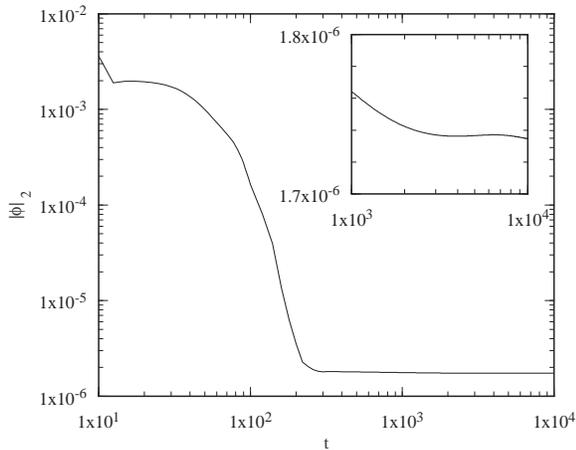}}
\caption{A log-log plot of the $\ell_2$ norm of the scalar field verses
time for a strong-field evolution.  The inset shows an expanded vertical
axis for the late time evolution.  The outer boundary is at $r=42M$.}
\label{fig:long}
\end{figure}

\begin{figure}
\centerline{\epsfxsize=3in\epsfbox{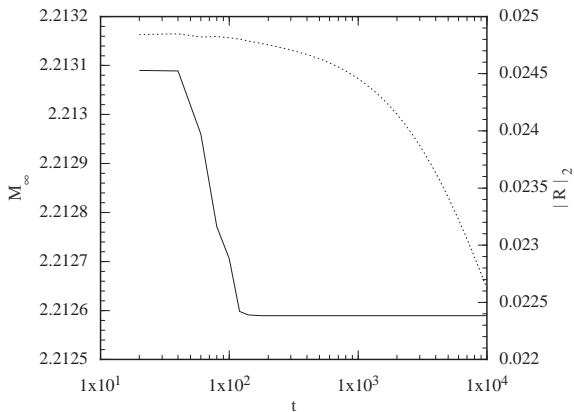}}
\caption{The mass (solid line) and the $\ell_2$ norm of the 4-Ricci scalar
(dotted line) verses time for a strong-field evolution.  The mass should be
nearly constant except for a small amount of scalar field which radiates to
infinity.  The plot of the Ricci scalar shows the drifting of the geometry
caused by the outer boundary conditions.  Contrary to the impression given
by this figure, the Ricci scalar is asymptoting to a fixed value.  The time
axis is logarithmic. }
\label{fig:longmr}
\end{figure}

Fig. \ref{fig:long} shows the $\ell_2$ norm of the scalar field during a
strong field evolution to $t=10000M$.  Fig. \ref{fig:longmr} shows the mass
and the Ricci scalar for the same evolution.  While the scalar field and
the mass fall off as expected, the plot of the Ricci scalar again shows the
``drifting'' of the geometry.

The plots of the Ricci scalar were made by discretizing to first order, the
analytic expression for $R$ derived in the usual way by $R=R^\mu{}_\mu$,
where $R_{\mu\nu}$ is the Ricci tensor.  Although the values of $R$ appear
large given that they are computed for a vacuum spacetime, they do converge
to zero to first order in the mesh spacing as expected.  Moreover, the
values of the individual additive terms in the expression for $R$ are
orders of magnitude larger than the scalar itself, indicating that $R$ is
the size we would expect for first order differencing at these resolutions.

\section{Mass Scaling}
\label{sec:mscale}
\begin{figure}
\centerline{\epsfxsize=3in\epsfbox{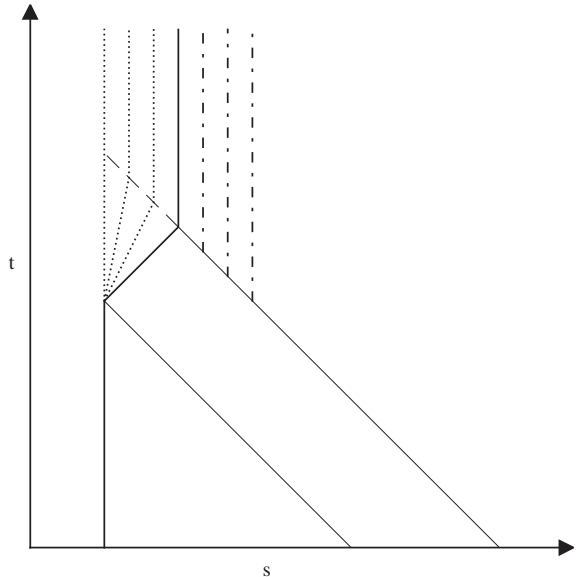}}
\caption{
Schematic motion of the horizon for various amplitudes of the scalar
field.  Notice that this diagram uses the areal coordinate $s$ and not the
radial coordinate $r$.  The solid dark vertical line which jogs right and
then continues vertically represents the critical path of the horizon.  The
dotted lines are sub-critical paths and the vertical dashed lines are
super-critical paths.  The two thin, diagonal lines represent the bounds of
the ingoing pulse of scalar field.  A super-critical pulse moves inward
until it crosses its gravitational radius.  Once this happens, the apparent
horizon jumps from its initial position to this new position where it
remains.  A sub-critical pulse moves inward until it encounters the
horizon.  If the field is very weak, the horizon is unaffected.  For
stronger fields, the horizon moves out until the pulse is entirely inside. 
For a critical pulse, the horizon moves out at the speed of light.  Note
however, that unless the energy density is a square wave, the horizon will
not move along the straight lines as shown in the diagram, but will move
along a curve with gradually increasing and then decreasing slope.
}
\label{fig:spacetime}
\end{figure}

As discussed in Sec.~\ref{sec:id} we focus study on the evolution of
initially ingoing ``Gaussian'' pulses of scalar radiation. In this case,
the infalling field exhibits two limiting behaviors, dependent on the
amplitude and width of the pulse.  These are: \emph{scattering} from the
existing black hole and \emph{collapse} to form a new horizon outside the
existing horizon.  Generically, these two behaviors are separated by a
\emph{critical} value of either amplitude or width.  Fig.
\ref{fig:spacetime} shows the path of the apparent horizon for various
amplitudes of initial data.

The final mass of the black hole should scale as a power of the amplitude
of the initial pulse.  To find out what this power should be, we can use
equation (\ref{eq:minf}).  Since the mass is conserved, $M_\infty$ is a
constant.  However, $M$ is not constant.  As the scalar field encounters
the horizon, some mass will be transferred from the integral term to $M$. 
The mass of the black hole will increase by an amount proportional to the
mass in the scalar field.  By this we mean that after the interaction we
have
\begin{equation}\label{eq:mincr}
M\to M+4\pi k \int^\infty_{r_h} s^2\left(\frac{\Phi^2+\Pi^2}{2a^2}+s
K^{\theta}{}_{\theta}\frac{\Phi\Pi}{a}\right) dr,
\end{equation}
where $k$ is a positive constant less than $1$.  For a very narrow pulse,
the entire mass of the field will go into the black hole, and hence, $k$
will be very close to $1$.  If the pulse if very wide, however, $k$ will be
very close to $0$.  Thus, to see how the final mass of the black hole
scales with the amplitude of the scalar pulse, we need only examine the
integral term in equation (\ref{eq:mincr}).

From equations (\ref{eq:phiinit})-(\ref{eq:Piinit}) we see that $\Phi$ and
$\Pi$ are proportional to $\phi$ and hence to $A$.  This means that the
integrand is proportional to $A^2$.  This of course assumes the dependence
of $a$, $K^{\theta}{}_{\theta}$, and $\beta$ on $\phi$ is much less than
the dependence of $m(r)$ on $\phi$.  While this seems a reasonable
assumption, it must be checked numerically in the strong field regime.

\begin{figure}
\centerline{\epsfxsize=3in\epsfbox{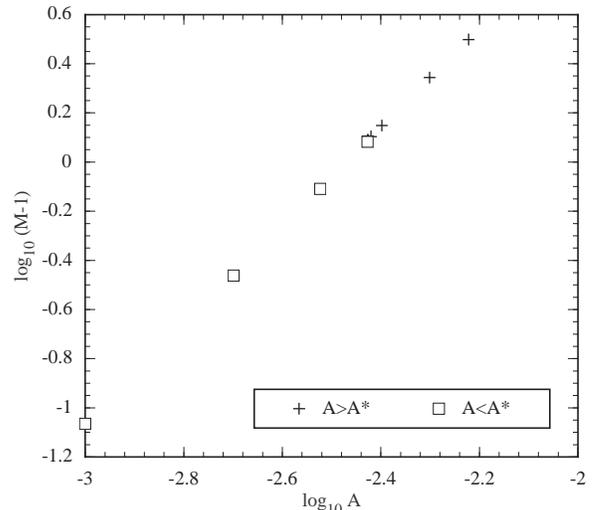}}
\caption{
A log-log plot of the final black hole mass verses the amplitude of the
scalar field pulse for initial data with $d=2$, $\sigma=2$, and $c=10$. The
squares represent data with amplitude less than the critical value, while
the crosses represent data with amplitude greater than the critical value. 
The slope of this line is $2.01$ showing that the black hole mass depends
on the square of the scalar field amplitude.
}
\label{fig:massd2all}
\end{figure}
\begin{figure}
\centerline{\epsfxsize=3in\epsfbox{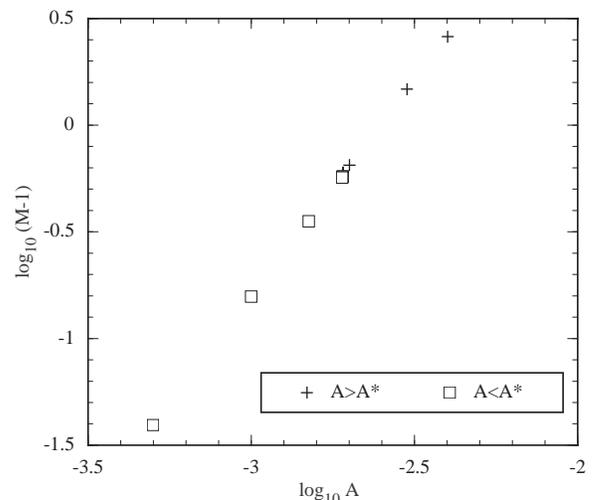}}
\caption{
A log-log plot of the final black hole mass verses the amplitude of the
scalar field pulse for initial data with $d=4$, $\sigma=2$, and $c=10$. 
The slope of this line is $1.99$ indicating again that the black hole mass
depends on the square of the scalar field amplitude.}
\label{fig:massd4all}
\end{figure}
Our numerical results verify this calculation.  For instance, the data in
Fig. \ref{fig:massd2all} is fit by the line
\begin{equation}\label{eq:msfit2}
\log\left( M-1\right) = 2.01\log A + 4.96,
\end{equation}
indicating that the mass grows with the square of the amplitude as
expected.  The graph also shows there is no difference in behavior for sub-
and super-critical data.  That is, the final mass of the black hole
exhibits the same dependence on the amplitude when the hole grows by
accretion as when it forms by collapse.

Similarly, the data in Fig. \ref{fig:massd4all} is fit by the line
\begin{equation}\label{eq:msfit4}
\log\left( M-1\right) = 1.99\log A + 5.19,
\end{equation}
indicating that the mass scaling is independent of the exact shape of the
ingoing pulse.

\section{Tails}
\begin{figure}
\centerline{\epsfxsize=3in\epsfbox{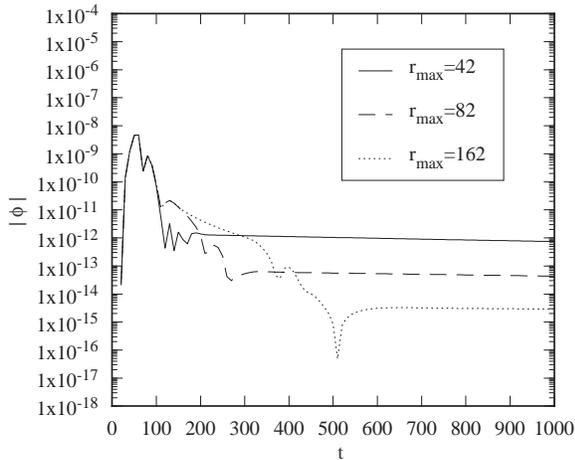}}
\caption{The absolute value of the scalar field at $r=30M$ verses time for
various spatial domains.  Notice the differences in late time fall off
caused by the different positions of the outer boundary.  The vertical
axis is logarithmic.}
\label{fig:tails}
\end{figure}

Fig. \ref{fig:tails} shows $\phi$ at constant $r$ for runs with
$r_{max}=42, 82, 162$.  It is clear that the position of the outer boundary
has a large effect on the late-time fall-off of the scalar field.  Even
with the sponge filter, there is enough reflection from the outer boundary
to cause the field to die off more slowly than it otherwise would.

\begin{figure}
\centerline{\epsfxsize=3in\epsfbox{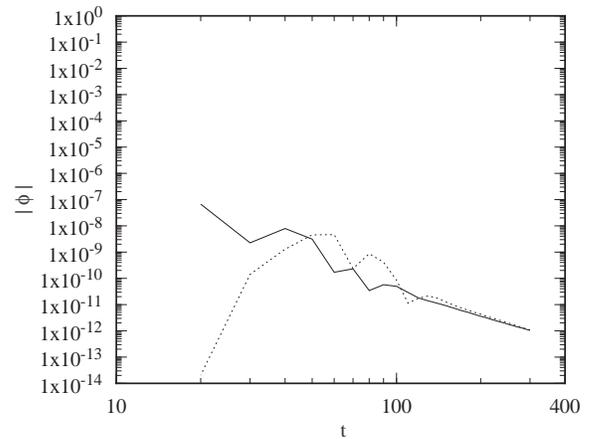}}
\caption{A log-log plot of the absolute value of the scalar field at the
horizon (solid line) and at $r=30M$ (dotted line) verses time for an
evolution with the outer boundary at $r=162M$.}
\label{fig:tail}
\end{figure}

However, with the outer boundary at $r_{max}=162$, it takes at least $300M$
for reflections from the scattered pulse to travel in from the outer
boundary and interfere with measurements at $r=30$, and still longer for
reflections to interfere with measurements at the horizon.  This should
provide enough time to accurately measure the rate of fall-off of the
scalar field.  Fig. \ref{fig:tail} shows the evolution of $\phi$ at $r=30$
and at the horizon up until $t=300M$.  A fit to the $r=30$ curve between
$200M$ and $300M$ shows $\phi$ falling off as $t^{-3.38}$.  A fit to the
horizon curve over the same range shows $\phi$ falling off as $t^{-3.06}$.
\label{sec:tails}

Linearized perturbation theory predicts these exponents should both be $-3$
\cite{gpp1994a}.  We note that previous evolutions carried out by 
Gundlach, Price, and Pullin gave exponents between $-2.63$ and $-2.74$ 
for $\phi$ at constant $r$ \cite{gpp1994b}.

\section{Ringing}
\label{sec:ring}
\begin{figure}
\centerline{\epsfxsize=3in\epsfbox{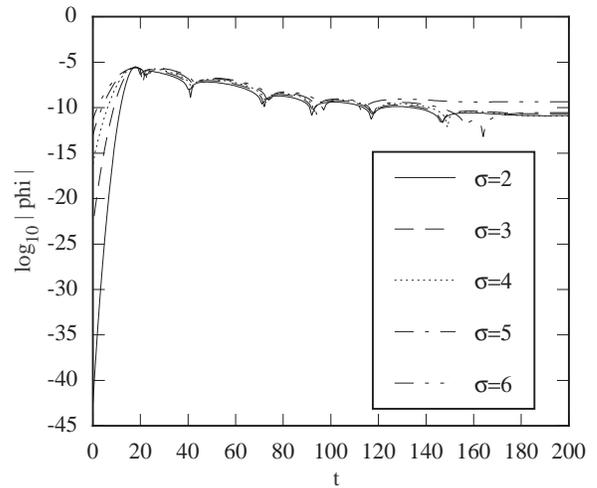}}
\caption{$\log |\phi|$ at the horizon verses time for various pulse widths
($A=2.0 \times 10^{-8}$).  While the curves match well for about $100M$, 
there are differences in late-time behavior exhibited by the wider pulses,
particularly those with $\sigma=5$ and $\sigma=6$.  This is an outer
boundary effect.  As the width becomes larger, less and less of the initial
pulse is absorbed by the black hole.  This means there is more scalar field
available to be reflected from the outer boundary and this causes
differences in the late-time evolution.
}
\label{fig:wphi1}
\end{figure}
\begin{figure}
\centerline{\epsfxsize=3in\epsfbox{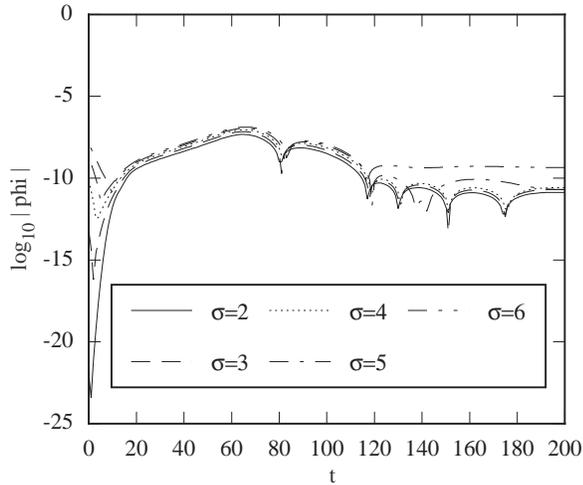}}
\caption{$\log |\phi|$ at $r=30$ verses time for various pulse widths
($A=2.0 \times 10^{-8}$).}
\label{fig:wphi2}
\end{figure}
\begin{figure}
\centerline{\epsfxsize=3in\epsfbox{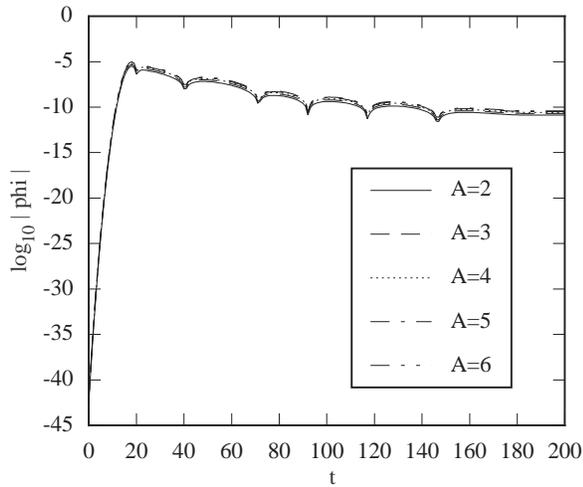}}
\caption{$\log |\phi|$ at the horizon verses time for various pulse
amplitudes ($\sigma=2.0$, amplitudes are $\times 10^{-8}$).}
\label{fig:wphi3}
\end{figure}
\begin{figure}
\centerline{\epsfxsize=3in\epsfbox{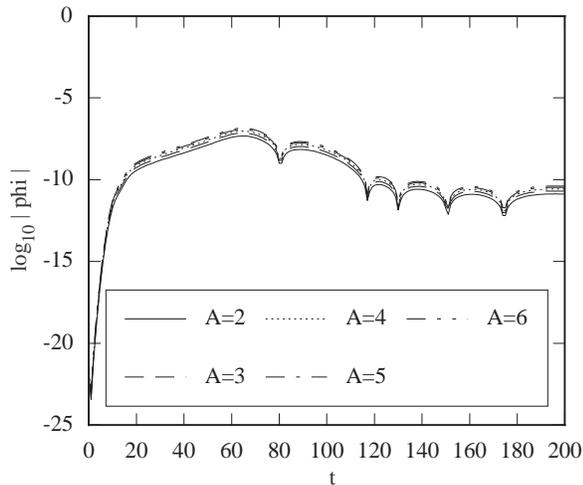}}
\caption{$\log |\phi|$ at $r=30$ verses time for various pulse amplitudes
($\sigma=2.0$, amplitudes are $\times 10^{-8}$).}
\label{fig:wphi4}
\end{figure}
Figs. \ref{fig:wphi1}-\ref{fig:wphi4} show the waveforms generated by the
scattering of packets of various widths and amplitudes.  The scalar field
is measured both at the horizon and at $r=30M$.  The oscillation period in
each of these figures is approximately $53M$, and is independent of initial
amplitude and pulse width.  Gundlach {\em et. al.} measured an oscillation
period of approximately $47M$ \cite{gpp1994b} during collapse of Gaussian
wave packets.

\section{Coordinate Effects}
\label{sec:ceff}
\begin{figure}
\centerline{\epsfxsize=3in\epsfbox{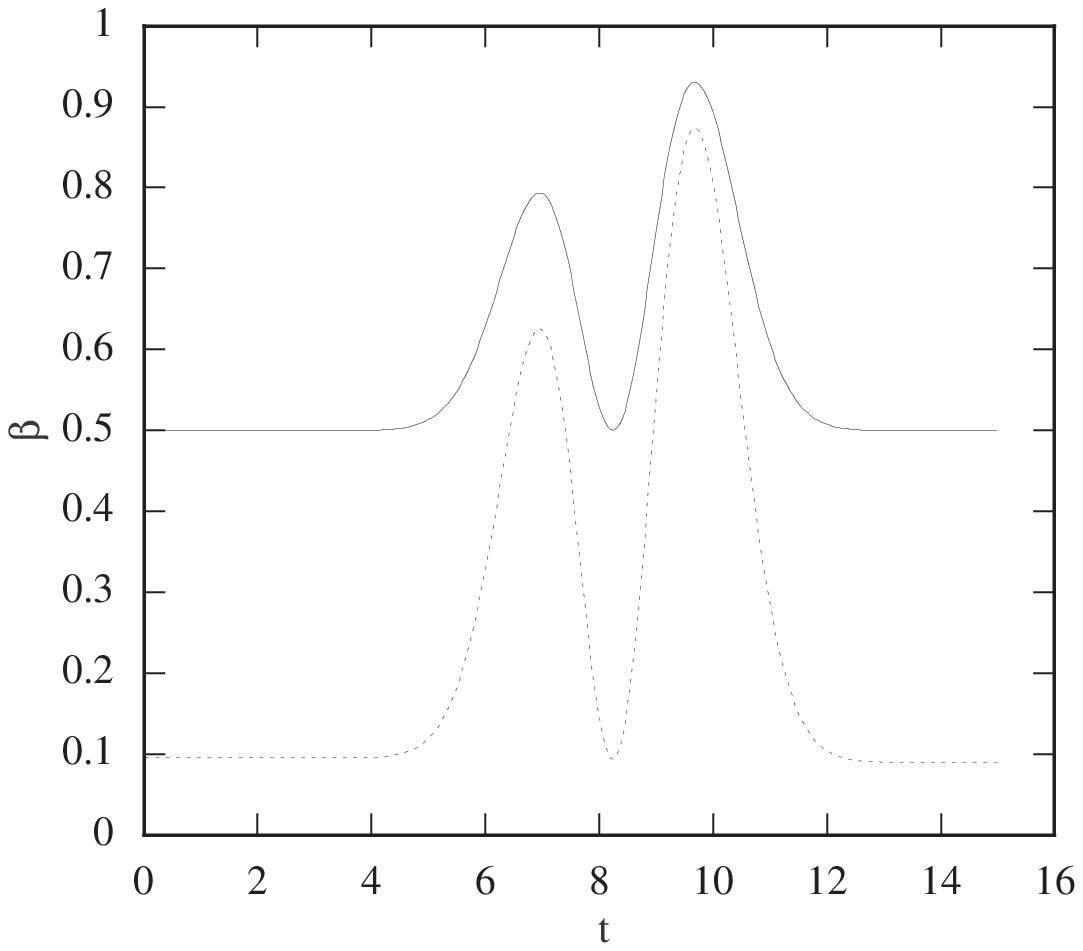}}
\caption{$\beta$ at the horizon (solid line) and the outer boundary 
(dotted line) for a near-critical solution with $\sigma=2$, $d=2$, and
$c=10$.  The outer boundary is at $r=42M$.  Outgoing pulses of scalar
radiation will exhibit retrograde motion whenever $\beta > .5$.}
\label{fig:betaio}
\end{figure}

Certain evolutions exhibit interesting effects which result from the use of
MMIEF coordinates.  Recall that the shift component, $\beta$
satisfies~(\ref{eq:betadef}).  At the apparent horizon,
(\ref{eq:mmiefaphor}) holds so we have 
\begin{equation}
\beta(r_h,t) = \frac{\dot{f(t)}}{2}+\frac{1}{2}.
\end{equation}
From this we can see that when no matter is crossing the
horizon, $\beta=1/2$ so the outgoing characteristic speed
(\ref{eq:charspeed}) is zero, as it
must be since the tracked surface is marginally trapped.  However, if
$\dot{f}=1$, then $\beta=1$ and the outgoing characteristic speed is $-1$. 
In this case, the light cone is degenerate.  In fact, from
(\ref{eq:betadef}) we can see that if $\dot{f}=1$, then $\beta=1$
\emph{everywhere}.  Does $\dot{f}$ ever equal one?  The most likely place
for this to happen is the critical solution because that is when the
``maximum'' amount of energy is crossing the horizon for a given pulse
shape.  The values of $\beta$ at the horizon and at the outer boundary
are plotted in Fig. \ref{fig:betaio} for a near-critical
solution.  $\beta$ gets as large as $.95$, but
never reaches 1. A near-critical solution from a family of  pulses with
$d=4$ gives a slightly higher maximum $\beta$, but still less than 1.  It
may be possible that a narrow enough pulse could cause $\dot{f}$ to reach 1
for an instant, but this has not been verified. Furthermore, it may be that
such a narrow pulse would collapse to form a new horizon before crossing
the existing horizon.
\begin{figure}
\centerline{\epsfxsize=3in\epsfbox{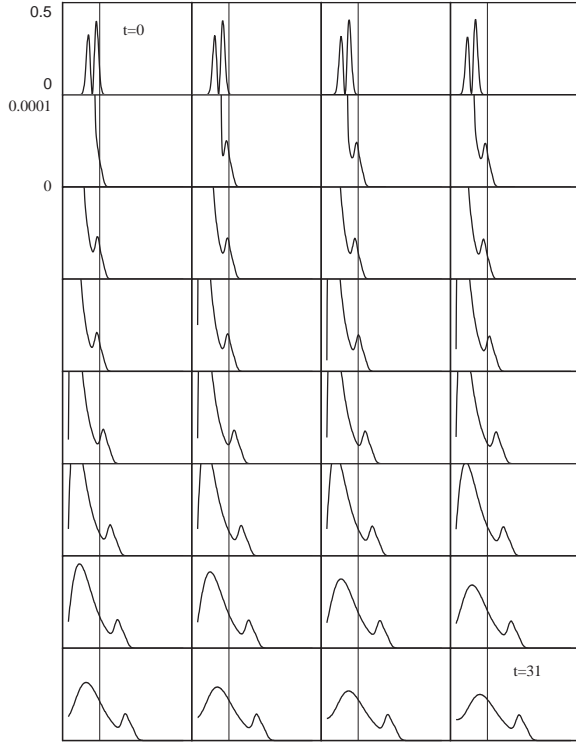}}
\caption{Evolution of $m'$ for a near-critical solution with
$\sigma=2$,$d=2$, and $c=10$.  Retrograde motion is apparent near $t=7M$
and $t=10M$.  The frames are spaced $1M$ apart in time.  The vertical scale
changes at $t=5M$ so that the small outgoing pulse can be observed.  The
thin vertical lines passing through the frames provide a common horizontal
reference to aid detection of the retrograde motion. }
\label{fig:mcdmdr}
\end{figure}

\begin{figure}
\centerline{\epsfxsize=3in\epsfbox{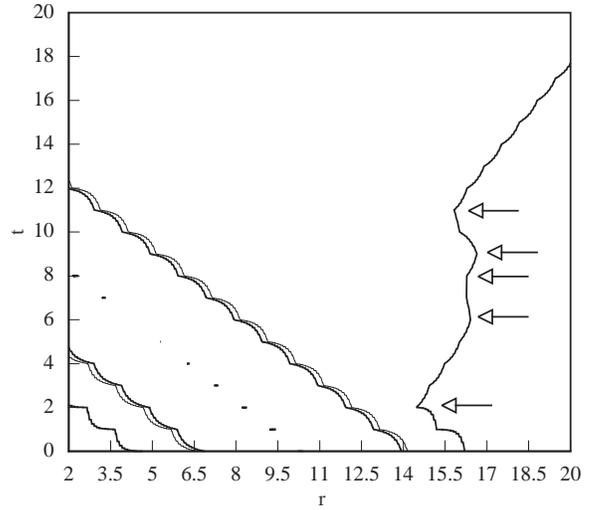}}
\caption{Contour plot of $m'$ for a near-critical solution with
$\sigma=2$,$d=2$, and $c=10$.  The small pulse moves out between the first
and second arrows then moves back in between the second and third arrows. 
It then moves out briefly between the third and fourth arrows and then
moves back in between the fourth and fifth arrows before renewing its
outward motion.
}
\label{fig:contour}
\end{figure}

\begin{figure}
\centerline{\epsfxsize=3in\epsfbox{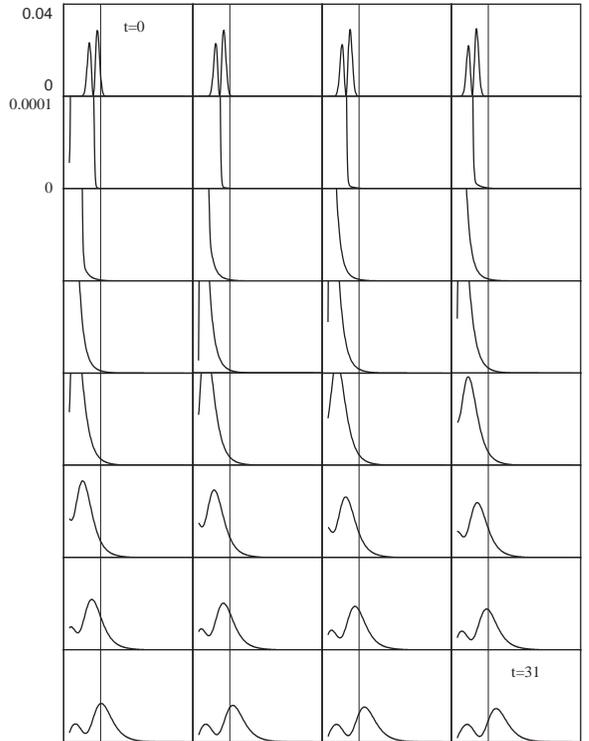}}
\caption{Evolution of $m'$ for a weak-field solution with
$A=.001$, $\sigma=2$,$d=2$, and $c=10$.  The frames are spaced $1M$ apart
in time.  The vertical scale changes at $t=5M$ the small outgoing pulses
can be observed.  There is no retrograde motion apparent.}
\label{fig:m6dmdr}
\end{figure}

Nevertheless, whenever $\beta > .5$, the outgoing characteristic speed is
negative. Thus, outgoing pulses will appear to move inward when plotted in
the radial coordinate, $r$.  Fig. \ref{fig:mcdmdr} shows an evolution of
$m'$ for the critical solution described above. There are two periods of
backwards motion; one at about $7M$ and the other at about $10M$.  These
are the times when each of the ``bumps'' crosses the horizon.  The
retrograde motion is easier to see in Fig. \ref{fig:contour} which shows
contours of $m'$ on a spacetime plot for the same evolution.  Fig.
\ref{fig:m6dmdr} shows a fairly weak field evolution of $m'$ for
comparison.  There is no retrograde motion in this case.

To completely remove these coordinate motion effects, we can abandon the
shifted areal coordinate $s$ and use the usual IEF coordinates.  In this
case, $r$ is areal again so $b=1$ instead of $1+f/r$.  The evolution and
constraint equations are trivially derived from equations
(\ref{eq:ham})-(\ref{eq:piev}) with the substitutions $s\to r$ and
$\dot{s}\to 0$.  Note that the function $f(t)$ no longer appears.
\begin{figure}
\centerline{\epsfxsize=3in\epsfbox{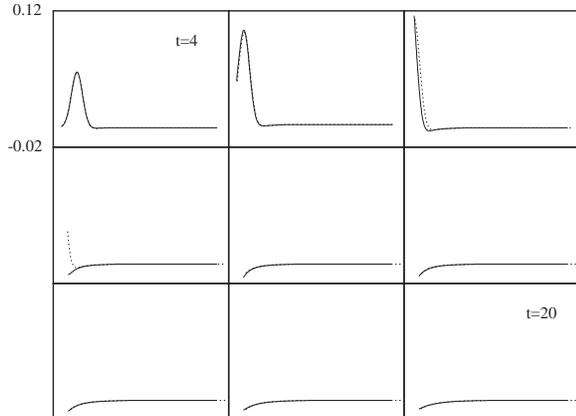}}
\caption{A comparison of $\phi$ computed in IEF coordinates (solid line)
compared with $\phi$ computed in MMIEF coordinates (dotted line).  The
frames are spaced $2M$ in time.  The horizontal axis extends from $0$ to
$45M$ in areal radius.  Differences between the two computations are caused
by small differences in the position of the inner boundary.}
\label{fig:iefcomp}
\end{figure}

Such a coordinate choice has the additional benefit of slightly simplifying
the equations.  However, the inner boundary of the grid will not be tied to
the apparent horizon as it is in MMIEF coordinates.  This causes no
problems in practice.  As matter crosses the horizon and the black hole
grows, some of the inner grid points are ``lost'' into the black hole and
we simply stop applying our difference equations there. Since the radial
coordinate is areal, the growth of the black hole is bounded by the total
mass of the spacetime, so there is no danger of having a large fraction of
the grid fall into the hole.  The previous one-dimensional black hole
excising calculations
\cite{seidelsuen1992,anninos1995,scheel1995a,scheel1995b} used
horizon-locked radial coordinates so that once the apparent horizon formed,
no further grid points would fall into the hole.  Thus, they should be
expected to exhibit similar coordinate motion effects to those seen in the
calculations using MMIEF coordinates. 

Fig. \ref{fig:iefcomp} shows a comparison of $\phi$ evolved in IEF
coordinates with $\phi$ evolved in MMIEF coordinates for a strong-field
case.  The two match closely with differences caused by small
differences in the positions of the inner boundary.

\section{Nonlinear Effects}
\label{sec:nonlin}
\begin{figure}
\centerline{\epsfxsize=3in\epsfbox{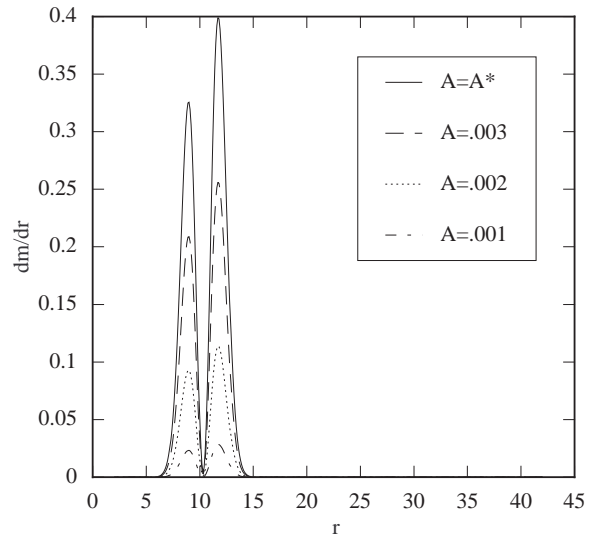}}
\caption{$m'$ at $t=0$ for $\sigma=2$, $d=2$, $c=10$, and various
amplitudes, including near-critical ($A^*$).}
\label{fig:dmdr10b}
\end{figure}
\begin{figure}
\centerline{\epsfxsize=3in\epsfbox{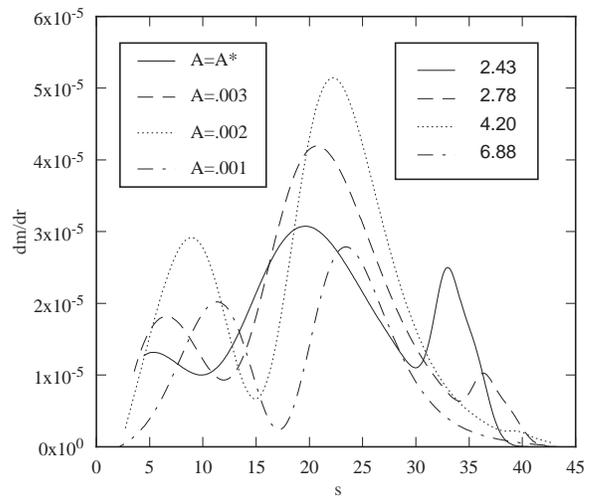}}
\caption{$m'$ at $t=40M$ for $\sigma=2$, $d=2$, $c=10$, and various
amplitudes.  The inset shows 1000 times fraction of mass scattered.  For
instance, the near-critical solution has $.243$ percent of its mass
scattered. }
\label{fig:dmdr10a}
\end{figure}
There is a sharp ``bump'' at the front of the outgoing pulse in Fig.
\ref{fig:mcdmdr}.  This feature is absent from the weak-field evolution of
Fig. \ref{fig:m6dmdr} and is certainly amplitude dependent.  Fig.
\ref{fig:dmdr10b} shows a series of initial pulse shapes for data with
various amplitudes all centered at $r=10$.  Fig.
\ref{fig:dmdr10a} shows the corresponding pulse shapes after scattering. 

\begin{figure}
\centerline{\epsfxsize=3in\epsfbox{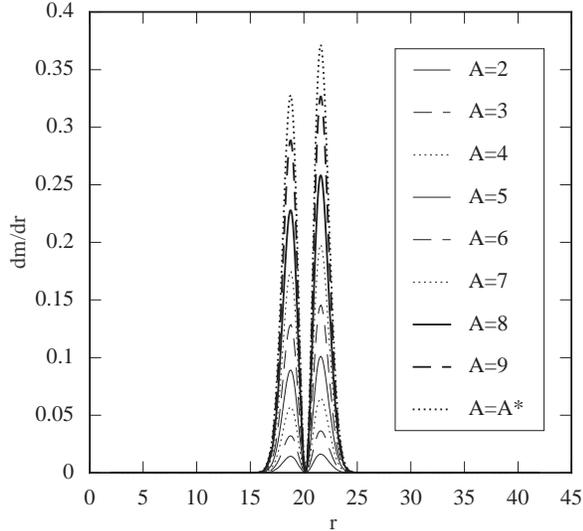}}
\caption{$m'$ at $t=0$ for $\sigma=2$, $d=2$, $c=20$, and various
amplitudes ($\times 10^{-4}$).}
\label{fig:dmdr20b}
\end{figure}

\begin{figure}
\centerline{\epsfxsize=3in\epsfbox{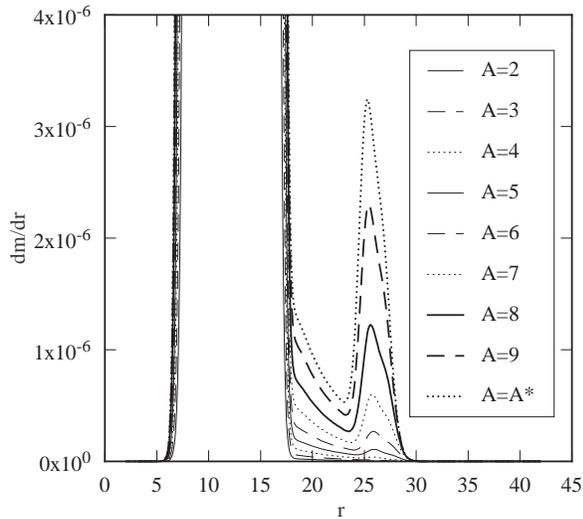}}
\caption{$M'$ at $t=8$ for $\sigma=2$, $d=2$, $c=20$, and various
amplitudes ($\times 10^{-4}$).  The small outgoing pulse appears before
scattering with the black hole.}
\label{fig:dmdr20a}
\end{figure}

Although this bump occurs at the front of the scattered pulse, it is not
caused by interactions with the black hole.  To see this more clearly, we
can start the pulse further out and see what happens.  Figs.
\ref{fig:dmdr20b} and \ref{fig:dmdr20a} show pulse shapes at $t=0$ and
$t=8$ for data with $c=20$. The outgoing bump develops for large amplitude
data without any help from the black hole.  A similar bump develops for
initial data with $d=4$.  These bumps always have the same characteristic
shape, though their widths may vary.

It is clear from Fig. \ref{fig:dmdr10a} that the amplitude of this outgoing
feature does not depend linearly on the initial amplitude of the pulse. For
instance, from Fig. \ref{fig:dmdr10b}, the height of the highest peak is
about 1.5 times the height of the next highest peak, while their widths are
the same.  However, the amplitude of the outgoing feature in the first case
is about 2.5 times the amplitude of the outgoing feature in the second
case, while their widths are the same.  Figures \ref{fig:dmdr20b} and
\ref{fig:dmdr20a} show similar nonlinear behavior for the pulses centered
at $r=20$.  It is difficult to tell whether this feature is caused by the
nonlinear interaction of an outgoing piece of the initial data with the
rest of the pulse, or if it is caused by back-scattering from the effective
self-potential of the ingoing pulse.  Further study is needed.

\section{Conclusions}
\label{sec:conc}
We have shown that the a \emph{null} based slicing works well with an
apparent horizon boundary condition, and results in a program which is
stable and convergent and will run forever.  Further, we have examined the
coordinate motion effects which result from this coordinate system and ways
to avoid them.  We have also seen an interesting nonlinear feature in the
self-gravitating scalar field.

While the calculations presented in this paper were carried out in
spherical symmetry, the coordinate system can be generalized to
multi-dimensions.  We have worked out an extension to the case of a Kerr
black hole in 3 dimensions (see \cite{marsa1995}), the details of which
will be presented elsewhere.

Unfortunately, this coordinate system can not be applied globally to
spacetimes which contain more than one black hole since it depends on the
notion of an areal coordinate.  Nevertheless, this technique is useful for
problems involving various matter fields and single black holes.  For
instance, it could be used to investigate the stability properties of the
solution found by Bechmann and Lechtenfeld \cite{bechmann1995} in their
investigation of the scalar no-hair theorem (see \cite{marsa1995} for work
in this direction).

\acknowledgments

This work was supported in part by the Texas Advanced Research Project
TARP-085 to Richard Matzner, the Cray Research Grant to Richard Matzner,
NSF PHY9318152 (ARPA supplemented) to the Grand Challenge Alliance, and NSF
PHY9510895 to the University of Pittsburgh.  Most of this work was done at
the University of Texas at Austin as part of \cite{marsa1995}.

It is MWC's pleasure to thank W.G. Unruh for his contributions
to the early stages of this research and particularly for the
suggestion that the Eddington-Finkelstein coordinate system
could be extended through the introduction of the shifted areal
coordinate, $s$.

\appendix
\section*{Spherically Symmetric Einstein-Klein-Gordon Equations}
\label{app:ssekge}

In spherical symmetry (using the usual spherical coordinate names
$(t,r,\theta,\phi)$, the three-metric $h_{ij}$ and the extrinsic curvature
tensor $K^i{}_j$ are diagonal.  We have 
\begin{equation}\label{eq:3met} 
h_{ij} = \hbox{{\rm diag}} \left( a^2 ( t, r) , r^2 b^2 (t,r), r^2 b^2
\sin^2 \theta \right) 
\end{equation}
\begin{equation}\label{eq:ecurv} 
K^i{}_j = \hbox{{\rm diag}} \left( K^r{}_r (t,r) , K^\theta{}_\theta (t,r),
K^\theta{}_\theta
\right) 
\end{equation}
\begin{equation}\label{eq:shift}
\beta^i = \left( \beta(t,r),0,0 \right) \quad
\alpha = \alpha \left( t,r \right) \quad
\phi = \phi(t,r)
\end{equation}
\begin{equation}\label{eq:spmet}
ds^2=\left(-\alpha^2+a^2\beta^2\right) dt^2 + 2 a^2\beta dt dr + a^2 dr^2 +
r^2b^2 d \Omega^2.
\end{equation}
The nonzero components of the Christoffel symbols are:
\begin{eqnarray}\label{eq:chr1}
\Gamma^r{}_{rr} = \frac{\partial_r a}{a} \quad \Gamma^r{}_{\theta\theta} = 
-\frac{rb\partial_r(rb)}{a^2} \quad \Gamma^{\theta}{}_{r\theta} =
\frac{\partial_r(rb)}{rb} \\
\Gamma^r{}_{\phi\phi} = -\sin^2 \theta \frac{rb\partial_r(rb)}{a^2} \quad
\Gamma^{\phi}{}_{r\phi} = \frac{\partial_r(rb)}{rb} \\
\Gamma^{\theta}{}_{\phi\phi}=-\sin \theta \cos \theta \quad
\Gamma^{\phi}{}_{\phi\theta} = - \cot \theta
\end{eqnarray}
The two non-zero components of the Ricci tensor are
\begin{equation}\label{eq:ricci1}
R^r{}_r = - \frac{2}{arb}\partial_r\frac{\partial_r(rb)}{a}
\end{equation}
\begin{equation}\label{eq:ricci2}
R^{\theta}{}_{\theta} = \frac{1}{ar^2b^2}\left[ a -
\partial_r\left(\frac{rb}{a}\partial_r\left( rb\right)\right)\right].
\end{equation}
The evolution equations for the metric components are
\begin{equation}\label{eq:speva}
\dot{a} = - a \alpha  K^r{}_r + \left( a\beta\right) '
\end{equation}
\begin{equation}\label{eq:spevb}
\dot{b} = - \alpha b  K^{\theta}{}_{\theta} + \frac{\beta}{r}\left(
r\beta\right)'.
\end{equation}
The evolution equations for the components of the extrinsic curvature are
\begin{eqnarray}\label{eq:spevkrr}
\dot{ K^r{}_r} = \beta  K^r{}_r ' &+& \alpha  K^r{}_r K -
\frac{1}{a}\left(\frac{\alpha '}{a}\right)' \nonumber \\ &-&
\frac{2\alpha}{arb}\left[\frac{(rb)'}{a}\right]' - \pi\alpha
\frac{\Phi^2}{a^2}
\end{eqnarray}
\begin{equation}\label{eq:spevktt}
\dot{ K^{\theta}{}_{\theta}} = \beta  K^{\theta}{}_{\theta} ' +
\alpha  K^{\theta}{}_{\theta} K + \frac{\alpha}{(rb)^2} -
\frac{1}{a(rb)^2} \left(\frac{\alpha r b}{a}(rb)'\right)'.
\end{equation}
The massless Klein-Gordon equation is formulated in terms of the two
auxiliary
fields $\Phi$ and $\Pi$ which are defined by
\begin{equation}\label{eq:defPhi}
\Phi \equiv \phi' \qquad
\Pi \equiv \frac{\alpha}{a}\left(\dot{\phi} - \beta \phi '\right)
\end{equation}
With these variables, the Klein-Gordon equation is
\begin{equation}\label{eq:spevPhi}
\dot{\Phi} = \left(\beta \Phi + \frac{\alpha}{a}\Pi\right)'
\end{equation}
\begin{equation}\label{eq:spevPi}
\dot{\Pi} = \frac{1}{r^2b^2}\left[ r^2b^2\left(\beta\Pi +
\frac{\alpha}{a}\Phi\right)\right]' - 2 \Pi \frac{\dot{b}}{b}.
\end{equation}
The Hamiltonian constraint is
\begin{eqnarray}\label{eq:spham}
&-& \frac{2}{arb}\left[\left(\frac{\left( rb\right)'}{a}\right)' +
\frac{1}{rb}\left(\left(\frac{rb}{a}\left(
rb\right)'\right)'-a\right)\right] \nonumber \\ &+& 4
K^r{}_r K^{\theta}{}_{\theta} + 2 K^{\theta}{}_{\theta}^2 =
8\pi\left(\frac{\Phi^2+\Pi^2}{a^2}\right) 
\end{eqnarray}
and the momentum constraint is
\begin{equation}\label{eq:spmom}
\frac{\left( rb\right)'}{rb}\left( K^{\theta}{}_{\theta} - K^r{}_r\right)-
K^{\theta}{}_{\theta}' =
-4\pi\frac{\Phi\Pi}{a}.
\end{equation}

\end{document}